\documentclass[pre,aps,floats,superscriptaddress,floatfix,twocolumn]{revtex4}
\usepackage{amssymb,amsmath}
\usepackage{amsmath,amssymb}
\usepackage{graphicx}
\usepackage{psfrag}
\usepackage{color}
\usepackage{dcolumn}
\usepackage[normalem]{ulem}

\def\beq{\begin{equation}}
\def\eeq{\end{equation}}
\def\bea{\begin{eqnarray}}
\def\eea{\end{eqnarray}}
\begin{document}
\title{Varieties of scaling regimes in hydromagnetic turbulence}
%  \title{What can scaling tell us about energy spectra in three-dimensional  
%fluid and magnetohydrodynamic turbulence?}
 \author{Abhik Basu}\email{abhik.basu@saha.ac.in,abhik.123@gmail.com}
\affiliation{Condensed Matter Physics Division, Saha Institute of
Nuclear Physics, Calcutta 700064, West Bengal, India}
\author{Jayanta K Bhattacharjee}\email{jayanta.bhattacharjee@gmail.com}
\affiliation{Department of Theoretical Physics, 2A and 2B Raja S C Mullick 
Road, Calcutta 700032, West Bengal, India}
\date{\today}
\begin{abstract}
We revisit the scaling properties of the energy spectra in fully developed incompressible
homogeneous turbulence in forced
magnetofluids (MHD) in three dimensions (3D), which are believed to be 
characterised 
by 
{\em universal 
scaling exponents} in the inertial range. Enumerating these universal scaling exponents 
that characterise the energy spectra remains a theoretical challenge.  To 
study this, we set up a scaling analysis
of the 3D MHD equations, driven by large-scale external forces and with or without a mean 
magnetic field. We use scaling arguments to bring out various scaling 
regimes for the energy spectra.  We obtain a variety of scaling in the inertial range,  ranging from the well-known Kolmogorov spectra 
in the isotropic 3D ordinary MHD to 
more complex scaling in the anisotropic cases  that depend on the magnitude of the mean magnetic field.  We further dwell on the possibility that the energy spectra scales as $k^{-2}$ in the inertial range, where $k$ is a wavevector belonging to the inertial range, and also speculate on unequal scaling by the kinetic and magnetic energy spectra in the inertial range of isotropic 3D ordinary MHD.
We predict the possibilities of
{\em scale-dependent anisotropy} and 
intriguing {\em weak dynamic scaling} in the Hall MHD and electron 
MHD regimes of anisotropic 
MHD turbulence. Our results can be tested in large scale simulations and 
relevant laboratory-based and solar wind experiments.
 
\end{abstract}

  \maketitle

  \section{Introduction}
  
    Kinetic energy spectrum in three-dimensional (3D) homogeneous and isotropic turbulence, described by the forced Navier-Stokes (NS) equation, displays 
universal scaling in the inertial range (that lies intermediate between the large forcing scales and small viscous dissipation scales) for sufficiently large Reynolds 
numbers~\cite{rahulrev}. For 
instance, the celebrated Kolmogorov dimensional 
analysis (hereafter K41)~\cite{k41} predicts that the kinetic energy 
spectrum $E_v({\bf 
k})\sim \langle |{\bf v}({\bf k},t)|^2\rangle k^2\sim k^{-5/3}$ (known as the 
K41 result in the literature~\cite{k41,frisch}) in the inertial range of the nonequilibrium turbulent steady 
states (NESS).  Here, $\bf k$ is a 
Fourier wavevector belonging to the inertial range and 
${\bf v}({\bf 
k},t)$ is the velocity field in the Fourier space; $\langle...\rangle$ refers to 
spatio-temporal averages in the NESS~\cite{multif}. 

  Magnetohydrodynamics (MHD) is the study of the 
properties of electrically 
conducting quasi-neutral fluids in the hydrodynamic limit, valid over huge 
ranges of spatial 
scales ranging from centimeters (e.g., laboratory plasmas) to very large scales 
in astrophysical settings (e.g., solar wind)~\cite{mont}. A plasma 
necessarily consists of two electrically charged components - ions and 
electrons. The dynamical equations for MHD depend crucially on the 
spatio-temporal scales of interests. For instance, at large spatial (scales 
larger 
than the ion Larmor radius) and temporal scales (times larger than 
$\omega_{pi}^{-1}$), both
the ion and electron motions are important, the local relative velocity between 
the 
electrons and the ions are negligible compared to the local center-of-mass 
velocity, for which the ordinary MHD description, which is a one fluid 
description, 
suffices~\cite{mont,jackson,arnab}. In a direct analogy with 
fluid dynamics, ordinary MHD can be viewed as a coupled dynamics of a 
velocity {\bf v} and a magnetic field $\bf b$, and the electric field drops out 
of the 
dynamics due to the condition of local charge neutrality.

Hall magnetohydrodynamics (HMHD) is again a  
single-fluid approximation that includes a Hall term in the Ohm's 
law (see below). This description extends
the validity domain of the ordinary MHD system to spatial scales down to a
fraction of the ion skin depth or frequencies comparable to the ion 
gyrofrequency~\cite{galtier}. More specifically, HMHD is a good description 
when we intend to describe the plasma dynamics up to
length scales shorter than the ion inertial length $d_i$ ($d_i = 
c/\omega_{pi}$, where $c$ is the speed
of light and $\omega_{pi}$ is the ion plasma frequency) and frequencies smaller 
than
than the ion cyclotron frequency  $\omega_{ci}$~\cite{galtier1}. For 
example, solar wind at small scales show signatures of HMHD~\cite{vkrishan}.
Eventually at sufficiently small scales $l<c/\omega_{pi}$ {\em and} at 
frequencies much higher than $\omega_{pi}$,
the ions are effectively frozen due to their larger inertia, and the electrons 
move in a frozen background of the ions, a regime aptly called electron MHD 
(EMHD)~\cite{emhd1,emhd2,cho}. EMHD phenomenology is believed to be operative 
in exotic astrophysical contexts like the crust of neutron 
stars~\cite{emhd-neutron}, solar corona and magnetotail~\cite{solar-corona} as 
well as laboratory experiments~\cite{lab-emhd}.
  
In equilibrium systems, fluctuations near a critical point (or a second order phase
transition) and in the ordered phases of systems with broken continuous symmetries show {\em universal dynamic scaling} in the long wavelength and long time scale limits~\cite{halpin}. Subsequently, the idea of universal scaling has been extended to nonequilibrium systems as well~\cite{ddlg,kpz,yakhot}. In fully developed fluid turbulence, this notion of universality implies that the scaling of the kinetic energy spectrum and the damping time-scale of the velocity fluctuations with
wavevectors in the inertial range is independent of the molecular viscosities~\cite{yakhot}.
  The question of scaling of energy 
spectra, both kinetic and magnetic, in 3D hydromagnetic 
 turbulence, although believed to be universal in the inertial range of
 fully developed MHD turbulence (i.e., in the 
large Reynolds number limit), is still not well-settled, either theoretically or 
experimentally.  
A major difference between 
3D homogeneous 
fluid turbulence and 3D homogeneous MHD turbulence originates from the possible 
presence of a mean magnetic field in magnetofluids: a mean fluid velocity, 
although makes the system nominally anisotropic, can be removed by a suitable 
Galilean transformation, thereby restoring full isotropy. In contrast, the 
magnetic field is invariant under Galilean transformations, and consequently, a 
mean magnetic field of magnitude $B_0$ cannot be removed by any Galilean boost 
and necessarily makes the system genuinely anisotropic. A non-zero $B_0$ 
introduces propagating modes in the form of Alfv\'en waves in ordinary 3D MHD  that 
have no 
analogues in homogeneous fluid turbulence. Similarly, in HMHD with 
$B_0\neq 0$, there are circularly polarised whistler and cyclotron 
modes~\cite{galtier} that are analogues of the Alfv\'en waves in ordinary 3D 
MHD, but have no counterparts in homogeneous and isotropic 3D fluid turbulence. The whistler modes
exist in anisotropic 3D EMHD as well~\cite{lyutikov}.
  
   Simple dimensional analysis similar to that for fluid turbulence suggests K41 scaling   in the inertial range for both the kinetic and magnetic energy spectra of 3D non-helical isotropic (i.e., no mean magnetic fields) MHD turbulence. However, some recent studies indicate the possibility of an unexpected $k^{-2}$ scaling of the energy spectra in the inertial range of 3D isotropic MHD~\cite{brand,anupam}. In addition, the presence of  mean magnetic fields, which gives rise to propagating 
  Alfv\'en waves, can significantly affect scaling.   In general,
  the effects of propagating waves on the scaling properties of driven systems 
 are still debated. MHD turbulence with a non-zero $B_0$ or Alfv\'en waves 
stands as a very good candidate to study this issue. In 
Refs.~\cite{verma,abjkb-jstat}, it has been argued within a low order 
perturbative 
analysis that the effective or renormalised mean magnetic field $B_{0R}$ 
(formally defined as the imaginary part of the field propagators 
at zero frequency) picks 
up singular corrections in the long wavelength limit. This in turn 
yields kinetic and magnetic spectra that are anisotropic in magnitude but 
display spatial scaling same as the K41 prediction. This prediction is 
different from the results from a 1D model for MHD turbulence~\cite{abjkb}, 
where the absence of any singular renormalisation of the mean magnetic fields 
render them irrelevant (in a scaling sense) in comparison with the viscous 
damping. This too, unexpectedly, yields the K41 scaling for the energy spectra. 
Numerical studies of Ref.~\cite{perez} revealed energy spectra closer 
to those predicted by the Iroshnikov-Kraichnan (IK) scaling of 
$k^{-3/2}$~\cite{ik}.  For large $B_0$, weak turbulence theories for incompressible MHD suggest a $k_\perp^{-2}$ scaling, where ${\bf k}_\perp$ is the component of the 3D wavevector $\bf k$, that is normal to the mean magnetic field~\cite{gal2}.
These multitude of predictions for scaling in 3DMHD calls for a generic scaling analysis for both isotropic and anisotropic 3D MHD. For HMHD 
and EMHD, there are predictions for scale breaking demarcating the long 
wavelength inertial range and an intermediate wavelength {\em dispersion 
range}~\cite{cho,phil-hall}. Recent 
experimental studies on table-top laser-plasma~\cite{nat-comm} reveal various 
scaling regimes at 
different ranges of wavevectors. In addition, very little is known 
about the dynamic scaling in MHD turbulence. Critical examination of  dynamic 
scaling regimes in MHD turbulence~\cite{dyn-mult} would be very useful as well.

In this article, we revisit the universal scaling of the kinetic and magnetic 
energy spectra in the inertial range in
3D turbulent homogeneous hydromagnetic fluids by employing scaling arguments. 
We cover (a) ordinary  3D MHD, (b) 3D HMHD and (c) 3D EMHD. We consider 
the role of a mean magnetic field in each of the above cases.  The scaling theory developed here
reveals a variety of scaling regimes. For instance, we find that (i) when forced at the
 largest scales and assuming non-helical MHD, the scaling of both the magnetic and kinetic spectra for 3D
isotropic ordinary MHD should follow the K41 prediction in the hydrodynamic long 
wavelength limit.  We further discuss the possibility of $k^{-2}$ for the magnetic spectrum; see  Ref.~\cite{brand}. This is obviously a ``weak scaling'' (where the magnetic and kinetic energy spectra scale differently), as made clear in Ref.~\cite{brand}, since the corresponding kinetic energy spectrum appears to scale very differently. The total energy, as mentioned in Ref.~\cite{brand} has a backward flux. Hence, unlike the strong scaling K41 spectra (where the magnetic and kinetic energy spectra scale identically), the forward cascade (i.e., from small to large wavenumbers) should be that of the other conserved quantity - the cross helicity. Furthermore, we speculate that $k^{-2}$ strong scaling can also be found in systems with large scale separations between the forcing scale and inertial range; see, e.g., this is possibly connected with
Ref.~\cite{anupam}. 
In contrast, with a 
finite $B_0$, i.e., with Alfv\'en waves present, the scaling generally takes an 
anisotropic form. Again with a large-scale forcing and neglecting helicity,  assuming the linear propagating Alfv\'en 
terms and the nonlinear terms scale in the same way for a finite $B_0$
the scaling of the energy spectra with ${\bf k}_\perp$ follows 
the K41 result, where as they scale differently with $k_\parallel$. This scaling behaviour in the limit of a very large $B_0$ that strongly suppresses the nonlinear terms, gives way to a $k_\perp^{-2}$ scaling~\cite{ng1}.  Here,  $k_\parallel$ is the component of the 3D wavevector $\bf k$  along ${\bf B}_0$. Further both 
the velocity and magnetic field fluctuations are characterised by the same 
dynamic exponent, corresponding to the more commonly found {\em strong 
dynamic scaling}~\cite{abmhdepl} (ii)  For 3D
HMHD, scale breaking at higher wavevectors are predicted together with 
anisotropic scaling of the energy spectra for a non-zero $B_0$. (iii) Lastly, 
the scaling of the magnetic energy spectrum in 3D EMHD is argued to be 
same as  3D HMHD but different 
  from 3D MHD. Scaling of the energy spectra in 3D anisotropic HMHD and EMHD are found to depend strongly on the magnitude of $B_0$. For instance, as $B_0$ rises, the scaling of the magnetic energy spectrum changes from $k_\perp^{-7/3}$ to $k_\perp^{-5/2}k_\parallel^{-1/2}$. We hightlight generic {\em scale-dependent} anisotropy in all regimes of 3D MHD for a non-zero $B_0$. We further predict the novel possibilities of unequal dynamic exponents for 
$\bf v$ and $\bf b$ fluctuations in a HMHD scaling regime when the Hall term in 
HMHD is dominant~\cite{weak}.  Occurrence of weak dynamic scaling is very rare in
natural systems. A prominent example is the equilibrium critical dynamics of symmetric binary mixture
near its demixing transition (a second order transition) point. Here,
the concentration fluctuations are {\em distinctly slower} than the velocity fluctuations,
reflecting the existence of two distinct dynamic exponents for the concentration and the
velocity~\cite{halpin,siggia,kawasaki,tauberbook}. It was also proposed that a model
for the ordered phase of the XY model also show similar weak dynamic scaling~\cite{peliti};
this was however ruled out later showing that at 3D there are no weak dynamic scaling
in this model~\cite{dohm}. More recently, a nonequilibrium version of Model C
is shown to display weak dynamic scaling for certain choices of the model parameters~\cite{uwe-pre}.
To our knowledge, 3D HMHD is the first candidate for weak dynamic scaling
in the realm of turbulence, which forms a principal prediction from the present study.
The rest of this article in organised as follows. In 
Sec.~\ref{mhd1} and Sec.~\ref{mhda1} we study scaling in ordinary isotropic 
and anisotropic 3D MHD respectively. Next we consider scaling 3D isotropic and 
anisotropic Hall MHD in Sec.~\ref{hmhd11} and Sec.~\ref{hmhdaniso} 
respectively. Finally, in Sec.~\ref{emhdsec} and Sec.~\ref{emhdaniso} analyse 
scaling in 3D isotropic and anisotropic EMHD respectively. In 
Sec.~\ref{summ} we summarise our results.

  \section{Scaling analysis}
  
  Scaling analysis is a powerful tool that is useful in extracting the dominant 
scaling 
behaviour in  the steady states of a dynamical system. In scaling analysis of a model, first space, time and 
the dynamical fields are scaled and next, the scale-invariance of the dynamical equations (invariance of the form of the dynamical equations under rescaling) for appropriate scaling factors for space, time and the dynamical fields is demanded.  For systems with uniform steady states, the dominant 
scaling behaviour in the steady state is ascertained by balancing the most relevant terms (in a scaling sense) in the long wavelength limit, and by imposing other conditions that characterise the steady states.

In order to set up the background, we first revisit 
scaling analysis of homogeneous and isotropic incompressible fluid turbulence that directly 
yields the K41 spectrum for the velocity field. Our discussions of scaling in this paper will be based on the premise that the governing equations, i.e., the evolution equations for $\bf v$ and $\bf b$ have to be invariant under a scale transformation that scales distances by $l$ and time by $l^{\tilde z}$. The {\em dynamic exponent} $\tilde z$ is an unknown which will be determined by some additional constraints. The additional constraint - a crucial ingredient in the scaling analysis - comes from the constancy (scale-independence) of the fluxes of the relevant conserved quantities in the ideal limit (i.e., in the absence of any external forcing or dissipation) at the intermediate scales or inertial range. When the dynamics is that of only one variable, e.g., velocity $\bf v$ for incompressible fluid turbulence, there can be no ambiguity. If there are more than one dynamical variables (e.g., two for incompressible 3D MHD) or more than one conserved quantities in the ideal limit (again as in 3D MHD), there is yet another additional issue about whether the fields will scale similarly or differently under the spatio-temporal rescaling. The former case turns out to be completely unambiguous. However, for the kind of scaling analysis that we carry out here, the latter cases are of ``if...then'' variety in some of the physical examples. It should also be noted that spatial anisotropy in the form of an externally imposed magnetic field will lead to the introduction of an additional scale, and the scaling arguments will hold under restrictive conditions which we will be able to specify.

%for hydrodynamic and hydromagnetic turbulence is  to enforce the scale-independence of the fluxes of certain quantities (which are conserved in the ideal or inviscid and unforced limit) in the 
%inertial range (see below). In a scaling analysis, such fluxes do not scale under any rescaling of the space, time and the dynamical variables. Imposing this together with balancing the dominant terms allow us to obtain the scaling behaviour of the system.

\subsection{ 3D fluid turbulence}\label{ns-scaling}

We revisit the universal scaling of the kinetic energy spectrum in 3D 
homogeneous and isotropic incompressible fluid turbulence. By using scaling arguments, we 
reproduce the well-known K41 result. 
The Navier Stokes equation for an incompressible velocity field $\bf v$ for an 
isotropic pure neutral fluid is given by~\cite{rahulrev,frisch}
 \begin{equation}
   \frac{\partial {\bf v}}{\partial t} + \lambda_1 ({\bf v}\cdot {\boldsymbol 
\nabla}){\bf v} = -{\boldsymbol\nabla} p + \nu \nabla^2 {\bf v} + {\bf 
f}_v,\label{ns1}
  \end{equation}
  together with the incompressibility condition given by 
${\boldsymbol\nabla}\cdot {\bf v}=0$. Here, $p$ and $\nu$ are the pressure and 
kinematic viscosity, respectively; ${\bf f}_v$ is a large-scale force needed to sustain 
fully developed turbulence. Parameter $\lambda_1$ takes the value unity, but 
is formally introduced in conventional renormalisation group based analysis 
for turbulence as a perturbative expansion parameter~\cite{yakhot}.  In the inviscid, unforced limit $(\nu =0, {\bf f}_v=0)$, Eq.~(\ref{ns1}) in 3D conserved the kinetic energy and fluid helicity. The kinetic 
energy spectrum in 3D is 
given by $E_v(k)\sim k^2 \langle |{\bf v ({\bf k,t}}|^2\rangle$ in the inertial range.  For a non-helical fluid turbulence, the kinetic energy flux in the steady state cascades from large length scales to small length scales and remains scale-independent in the intermediate inertial range. The physical argument behind this is the fact that energy is injected from outside
at the largest (forcing) scales (by the large scale external forces) and get dissipated at very small scales by the molecular viscosities (viscous scales). In the intervening inertial regime in the steady state, the energy 
just flows from the large scales to small scales, without any energy injection or dissipation. This
ensures that the energy flux is constant in the inertial range~\cite{flux1}. The 
well-known Kolmogorov dimensional analysis predicts $E_v (k)\sim k^{-5/3}$~\cite{k41,frisch}. We 
will see below how this result may be recovered from a simple scaling analysis.

To begin with we scale space $\bf x$ time $t$ and 3D velocity $\bf v$ as 
follows:
\begin{equation}
 {\bf x}\rightarrow l {\bf x},\,\,t\rightarrow l^{\tilde z} t,\,\, {\bf v} 
\rightarrow l^a {\bf v},\label{resc1}
\end{equation}
where $\tilde z$ is the dynamic exponent. 
Demanding scale invariance, we obtain (in a scaling sense)
\begin{equation}
 \frac{\partial {\bf v}}{\partial t}\sim {\bf v}\cdot {\boldsymbol\nabla} {\bf 
v}\implies l^{a-\tilde z}=l^{2a-1}\implies a=1-\tilde z,\label{cond1}
\end{equation}
that is consistent with the physical dimension of a velocity.  We note that the 3D NS equation (\ref{ns1}) in the inviscid limit ($\nu =0$), 
or the Euler equation is scale-invariant, i.e., its form remains unchanged, with $a=1-\tilde z$ 
for arbitrary $\tilde z$. For $\nu>0$, this symmetry gets restricted as we show below. Here, we have 
used that the nonlinear coupling $\lambda_1$ {\em does not} pick up any 
scale-dependences under rescaling (\ref{resc1}) that is consistent with its 
nonrenormalisation due to the Galilean invariance of Eq.~(\ref{ns1}). Viscosity 
$\nu$ is assumed to pick a scale-dependence that is consistent with the value 
of $\tilde z$ (obtained below). 

In a mean-field like approach, the kinetic energy flux or the kinetic energy dissipation rate per unit mass $\epsilon_v$, neglecting intermittency, is assumed to be a constant in the inertial range of the steady states of fully developed fluid turbulence~\cite{frisch}, and should not change under rescaling (\ref{resc1}).
Thence, demanding scale-independence 
of $\epsilon_v$ we find:
\begin{equation}
\epsilon_v\sim \frac{\partial v^2}{\partial t} \sim l^0\implies 2a = \tilde z. \label{cond2}
\end{equation}
Combining (\ref{cond1}) and (\ref{cond2}), we find
\begin{equation}
 a=\frac{1}{3},\;\;\tilde z=\frac{2}{3},
\end{equation}
which are in agreement with the results obtained in Ref.~\cite{yakhot}.

Next step is to calculate the scaling of the kinetic energy spectrum from the 
values of $a$ and $\tilde z$, already known as above. We start by noting that
\begin{equation}
 \langle {\bf v}({\bf k},t)\cdot {\bf v}({\bf k'},t)\rangle=F_v (k)\delta ({\bf 
k+k'}),\label{evk}
\end{equation}
where $F_v(k)$ is related to $E_v(k)$ (see below).
%Here $\bf k$ is a Fourier wavevector.
Noting that
\begin{equation}
 {\bf v}({\bf k},t)\sim \int d^3x \exp (-i{\bf k\cdot x}) {\bf v}({\bf x},t), 
\label{vfour}
\end{equation}
under rescaling (\ref{resc1}) we have
\begin{equation}
 {\bf v}({\bf k},t)\sim l^{a+3}\sim k^{-a-3},
\end{equation}
where $k\sim l^{-1}$, in a scaling sense. Next, equating the scale factors on 
both sides
of (\ref{evk}), we obtain
\begin{equation}
 F_v(k)\sim k^{-3-2a}.\label{evk1}
\end{equation}
Now, the kinetic energy 
spectrum $E_v(k)$ in 3D is given by
\begin{equation}
 E_v(k)\sim k^2 F_v(k)\sim k^{-1-2a}\sim k^{-5/3},
\end{equation}
in agreement with Ref.~\cite{yakhot}. Notice that $\tilde z=2/3$ together 
with a scale-independent kinetic energy flux implies the effective kinematic viscosity 
scales as $\nu l^{2-z}\sim\nu l^{4/3}$~\cite{yakhot} that control the relaxation of 
the $\bf 
v$-fluctuations restores the scale-invariance of (\ref{ns1}). That the 
effective viscosity should be scale-dependent in order to keep the kinetic 
energy flux scale-independent has been known ever since the seminal works by 
Heisenberg~\cite{heisenberg} and Chandrasekhar~\cite{chandra}. This opens up the distinct possibility that in systems with more than one dynamical variables and independent fluxes, more than one 
dynamic exponents may be needed to keep the fluxes scale-independent.

\subsection{Ordinary 3D MHD turbulence}
Here we first consider homogeneous and isotropic incompressible 3D MHD turbulence, followed by 
its anisotropic analogue.

\subsubsection{Isotropic 3D MHD turbulence}
\label{mhd1}

  The ordinary 3DMHD equations for an incompressible homogeneous 
  and isotropic magnetofluid are composed of 
the generalised Navier-Stokes equation for the velocity field $\bf v$ and 
Induction equation for the magnetic field $\bf b$~\cite{jackson,arnab}. These are, respectively,
  \begin{equation}
   \frac{\partial {\bf v}}{\partial t} + \lambda_1 ({\bf v}\cdot {\boldsymbol 
\nabla}){\bf v} = -{\boldsymbol\nabla} p + \lambda_2 ({\bf 
b}\cdot{\boldsymbol\nabla}) {\bf b} + 
\nu \nabla^2 {\bf v} + {\bf f}_v, \label{eqmhd1}
  \end{equation}
  and
\begin{equation}
 \frac{\partial {\bf b}}{\partial t} + \lambda_1({\bf v}\cdot 
{\boldsymbol\nabla}) {\bf 
b} = \lambda_1({\bf b}\cdot {\boldsymbol\nabla}) {\bf v} 
+ \mu\nabla^2 {\bf b} + {\bf f}_b. \label{eqmhd2}
\end{equation}
 The 
effective pressure $p$ now includes the magnetic contribution $b^2/2$. 
Furthermore, $\lambda_1,\lambda_2$ are nonlinear coupling constants. 
Parameters $\nu$ and $\mu$ are kinematic and magnetic viscosities. We impose 
incompressibility ${\boldsymbol\nabla}\cdot {\bf v}=0$ and 
${\boldsymbol\nabla}\cdot {\bf b}=0$. Functions ${\bf f}_v$ and ${\bf f}_b$ are 
external  stochastic forces. As for (\ref{ns1}) $\lambda_1=1$ and $\lambda_2$ 
just sets the scale of $\bf b$ 
with respect to $\bf v$~\cite{verma,abjkb-jstat}. Similar to Eq.~(\ref{ns1}), 
Eqs.~(\ref{eqmhd1}) and (\ref{eqmhd2})
are invariant under Galilean transformation~\cite{verma,abjkb-jstat}) that
ensures nonrenormalisation
of $\lambda_1$ in a RG framework. Further, as pointed out in 
Ref.~\cite{verma,abjkb-jstat},
working in terms of {\em effective} magnetic fields that leaves Eq.~(\ref{eqmhd2})
unchanged, leads to nonrenormalisation of $\lambda_2$ as well.   Hence, without any loss of generality, we set $\lambda_1=\lambda_2=1$ in what follows below. The absence of any
mean magnetic field implies that $\langle {\bf b}({\bf x},t)\rangle =0$. 
Functions ${\bf f}_v$ and ${\bf f}_b$ are external large scale forces need to 
maintain fully developed MHD turbulence.  Equations~(\ref{eqmhd1}) and (\ref{eqmhd2}) in the inviscid, unforced limit in 3D conserve the total energy $E=\int_x (v^2 + b^2)$, cross helicity $H_c= \int_x {\bf v}\cdot {\bf b}$ and the magnetic helicity $H_m=\int_x {\bf A}\cdot {\bf b}$, where ${\bf A}$ is the vector potential for $\bf b$: ${\bf b}={\boldsymbol\nabla}\times {\bf A}$.

The scaling of the kinetic and magnetic spectra in the inertial range
can be easily obtained by generalising the analysis developed in 
Sec.~\ref{ns-scaling}.
Scaling ans\"atze (\ref{resc1}) is now to be augmented by the scaling
of $\bf b$:
\begin{equation}
 {\bf b}\rightarrow l^y {\bf b}.\label{bsc}
\end{equation}
As before, we demand scale invariance of Eqs.~(\ref{eqmhd1}) and (\ref{eqmhd2}). 
We consider large-scale forcings and assume non-helical MHD turbulence, i.e., $H_c\approx0,\,H_M\approx 0$. Now balancing the nonlinear terms in (\ref{eqmhd1}) % with $\partial {\bf b}/\partial t$
we obtain (in a scaling sense) 
\begin{equation}
 a=y.
\end{equation}
Notice that with $\tilde z=1-a$, the nonlinear terms in Eq.~(\ref{eqmhd2}) scale in the same way as $\partial {\bf b}/\partial t$:
\begin{equation}
\frac{\partial {\bf b}}{\partial t}\sim ({\bf v}\cdot {\boldsymbol\nabla}) {\bf b} \sim ({\bf b}\cdot {\boldsymbol\nabla}) {\bf v}.
\end{equation}

Due to the equality $a=y$, scale-independence of the kinetic (magnetic) 
energy flux automatically ensures scale-independence of the magnetic (kinetic) 
energy flux.  This then corresponds to scale-independence of the total energy flux. Proceeding as in Sec.~\ref{ns-scaling}, we then find
\begin{equation}
 2a=2y=\tilde z,
\end{equation}
giving $a=y=1/3$ and $\tilde z=2/3$. We can now proceed to obtain the scaling 
of the both kinetic and magnetic energy spectra in the inertial ranges by 
following the logic outlined in Sec.~\ref{ns-scaling} above. Similar to ${\bf 
v}({\bf k},t)$ we define ${\bf b} ({\bf k},t)$ via
\begin{equation}
 {\bf b}({\bf k},t) \sim \int d^3 x \exp (-i{\bf k\cdot x}) {\bf b}({\bf x},t), 
\label{bfour}
\end{equation}
yielding 
\begin{equation}
 {\bf b}({\bf k},t)\sim l^{y+3}\sim k^{-y-3}.
\end{equation}
Analogous to (\ref{evk}) we further define
\begin{equation}
 \langle {\bf b}({\bf k},t)\cdot {\bf b}({\bf k'},t)\rangle=F_b (k)\delta ({\bf 
k+k'}),\label{ebk}
\end{equation}
yielding as for $F_v(k)$
\begin{equation}
 F_b(k)\sim k^{-3-2y}\label{ebk1}.
\end{equation}
Thus, the magnetic energy 
spectrum $E_b(k)$ in 3D  scales as
\begin{equation}
 E_b(k)\sim k^2 F_b(k)\sim k^{-1-2y}\sim k^{-5/3}.
\end{equation}
The scaling of $E_v(k)$ remains unchanged from what we obtained in 
Sec.~\ref{ns-scaling}. Lastly, $\tilde z=2/3$ indicates that both $\nu$ and 
$\mu$ scale as $l^{2-\tilde z}$ in the inertial range. Thus, both $\bf v$ and 
$\bf b$-fluctuations are characterised by the same $\tilde z$, or strong 
dynamic scaling prevails. That both $\bf v$ and $\bf b$ must have the same 
$\tilde z$, can also be argued the scale-dependences of effective $\nu$ and 
$\mu$ needed to make the magnetic and kinetic energy spectra must be the same.  That we find $a=y$ in non-helical isotropic 3D MHD is consistent with the discussions in Ref.~\cite{anupam}.

 So far we have considered scale-independence of only the energy flux (kinetic and magnetic), which straight forwardly leads to K41 scaling for  both the energy spectra. This is justified when the total cross helicity and magnetic helicity are zero.  Can $E_v(k)$ and $E_b(k)$ ever display non-K41 type inertial range scaling in any situation? Recent studies in Refs.~\cite{brand,anupam} suggest that even in 3D isotropic MHD turbulence, the energy spectra can be non-K41 type; Ref.~\cite{brand} found $k^{-2}$ scaling where as Ref.~\cite{anupam} found both $k^{-2}$ and IK spectra, in addition to K41 spectra. We now discuss possible ways  to generalise the scaling theory to allow for non-K41 type inertial range scaling by $E_v(k)$ and $E_b(k)$. This can be then used to study the results in Refs.~\cite{brand,anupam}.  In Ref.~\cite{anupam} bounds on the scaling exponents of $E_+(k)\sim k^{q_+}$ and $E_-(k)\sim k^{q_-}$ in the inertial range are discussed, where $E_\pm$ are the energy spectra of the Els\"asser variables, which are just linear combinations of ${\bf v}$ and $\bf b$. It has been argued that in the absence of any cross-helicity, $q^+=q^-\neq 3/2$. This would be necessarily mean $a=y$ in our notation and $E_v(k)\sim E_b(k)$. The solution $a=y=1/3$ corresponding to the K41 spectra satisfy this. However, it is also known that if there is large scale-separation between the forcing scale and the inertial range then effective anisotropy in the inertial range can emerge and the magnetic fields  in the forcing scale can play the role of background magnetic fields for the fluctuating magnetic fields in the inertial range~\cite{anupam}. This should naturally generate Alfv\'en wave-like excitations with linear dispersion (see also below).  K41 scaling follows when this is subdominant to the nonlinear cascade.   In contrast, when this dominates over the nonlinear interactions in the inertial range,  $\tilde z=1$. If we further impose scale-independence of the kinetic and magnetic energy fluxes, then we find $a=y=\tilde z/2=1$. Following the logic outlined above, this yields
\begin{equation}
 E_v(k)\sim E_b(k)\sim k^{-2},
\end{equation}
see, e.g., Refs.~\cite{brand,anupam}. Furthermore, Ref.~\cite{anupam} has argued than in the presence of cross-helicity $q_+\neq q_-$, equivalently, $a\neq y$ is possible. The emergence of unequal scaling of $v$ and $b$ effectively implies the existence of an additional dimensional parameter as in the {\em incomplete self-similarity} discussed in Ref.~\cite{cono}. We construct such a possibility below. At the outset, we assume $a=y$ and set $y=a+\alpha$, where $\alpha\neq 0$ implies scale-breaking. Assuming there is no anomalous scaling of $\bf v$, $a=1-\tilde z$, consistent with the dimension of $\bf v$. Further, assume the cross-helicity flux $\epsilon_c$ to be the only relevant (forward) flux in the problem. Demanding scale-independence of $\epsilon_c$, we obtain
\begin{equation}
 2a+\alpha=\tilde z.
\end{equation}
Magnetic energy spectrum $E_b(k)\sim k^{-1-2y}\sim k^{-1-2a-2\alpha}$. Thus, $a+\alpha=1/2$ would give $E_b(k)\sim k^{-2}$. Together with the conditions on $a,\alpha$ and $\tilde z$, this implies 
$\alpha=1/4$. Of course, for other values of $\alpha$, the scaling of $E_b(k)$ will change. The scaling analysis cannot however precisely evaluate the scaling exponents $a,\alpha$ and $\tilde z$.
We note that the magnetic energy per unit volume $V$
\begin{equation}
 \frac{1}{V} \int \, b^2 ({\bf x})\,d^3x \propto \int \,dk k^2 \langle |{\bf b}({\bf k})|^2\rangle \equiv \int B(k) dk.
\end{equation}
Therefore, on dimensional ground
\begin{equation}
 B(k)\sim l^{3+2\alpha -2\tilde z}.\label{brandeq1}
\end{equation}
Assuming $B(k)$ is to be constructed from the cross-helicity flux $\epsilon_c\sim \partial ({\bf v\cdot b})/\partial t$, we write 
\begin{equation}
 B(k)\sim \left[\frac{vb}{t}\right]^\beta l^\gamma \label{brandeq2}
\end{equation}
for arbitrary $\tilde z$ on dimensional ground. Now comparing (\ref{brandeq1}) and (\ref{brandeq2}), we find $\beta =2/3$ and 
\begin{equation}
 \gamma=\frac{5}{3} + \frac{4\alpha}{3}.
\end{equation}
The weak turbulence scaling exponent $\gamma=2$ (i.e., $E_b(k)\sim k^{-2}$) is obtained for $\alpha =1/4$, as we have found above. In addition, we find $E_b(k)$ scales as $\epsilon_c^{2/3}$, a result that can be tested in directed numerical simulations of 3D MHD equations. The scaling of $E_v(k)$ will be different from $E_b(k)$. We do not comment on that here.  Thus the generalised scaling theory indeed predicts $k^{-2}$ scaling by $E_b(k)$ as one possible solution for scaling, but does not rule other scaling solutions, and more interestingly, generally allows for different scaling by $E_b(k)$ and $E_v(k)$.

\subsubsection{Alfv\'en waves: effects of anisotropy on scaling}\label{mhda1}

Most natural realisations of a plasma usually contain a mean magnetic field, 
e.g., tokamak plasma and solar wind~\cite{solar}.
Thus it is pertinent to consider now how a mean magnetic field can alter the 
scaling behaviour
elucidated above. We choose the mean magnetic field ${\bf B}_0$ to be along the 
$z$-axis, 
which lead to additional linear terms in Eqs.~(\ref{eqmhd1}) and 
(\ref{eqmhd2}). %We also assume $B_0 \gg b$, typical magnetic field fluctuations.
\begin{equation}
   \frac{\partial {\bf v}}{\partial t} +  ({\bf v}\cdot {\boldsymbol 
\nabla}){\bf v} = -{\boldsymbol\nabla} p +  ({\bf 
b}\cdot{\boldsymbol\nabla}) {\bf b} + B_0 \frac{\partial {\bf b}}{\partial z} + 
\nu \nabla^2 {\bf v} + {\bf f}_v, \label{eqmhd1a}
  \end{equation}
  and
\begin{equation}
 \frac{\partial {\bf b}}{\partial t} + ({\bf v}\cdot 
{\boldsymbol\nabla}) {\bf 
b} = ({\bf b}\cdot {\boldsymbol\nabla}) {\bf v} 
+ B_0 \frac{\partial {\bf v}}{\partial z}
+ \mu\nabla^2 {\bf b} + {\bf f}_b. \label{eqmhd2a}
\end{equation}
The linear terms in Eqs.~(\ref{eqmhd1a}) and (\ref{eqmhd2a}) allow for 
underdamped propagating waves, known as the {\em Alfv\'en waves}~\cite{mont} 
in the literature, with a dispersion
\begin{equation}
 \omega \propto B_0 k_\parallel + O(k^2), \label{alfdisp}
\end{equation}
in the long wavelength limit, where $\omega$ is a Fourier frequency and $k_\parallel$ 
is the $z$-component of $\bf k$; ${\bf k}=({\bf k}_\perp,k_\parallel), {\bf k}_\perp = (k_x,k_y)$.  We continue to assume large-scale forcings and the system to have negligible helicity. %For reasons identical to the corresponding 
%isotropic case discussed above, coupling constants $\lambda_1$ and $\lambda_2$ 
%do not renormalise. 
 
 Noting that a non-zero $B_0$ necessarily makes the system anisotropic, we need 
to generalise the scaling ans\"atze to account for anisotropy. In particular, 
we now expect spatially anisotropic scaling with the length scales in the $xy$ 
plane to scale different from those along the $z$-axis.  Without any loss of generality, we set 
\begin{equation}
{\bf x}\rightarrow l_\perp {\bf 
x}, {z}\rightarrow l_\parallel { z},\, t\rightarrow l_\perp^{\tilde z},\,{\bf 
v}\rightarrow l_\perp^a{\bf v},\,{\bf b}\rightarrow l_\perp^y {\bf b}, 
\label{resc-ani}
\end{equation}
where $l_\perp$ is a length scale in the $xy$-plane~\cite{foot2}.
We 
further set length scale along the $z$-axis $l_\parallel \sim l_\perp^\xi$ that 
controls the relative scaling 
between the $xy$-plane and the $z$-axis; for $\xi\neq 1$, the system is 
anisotropic. Here ${\bf x}=(x,y)$ is the in-plane coordinate. We also define 
${\boldsymbol\nabla}_\perp=(\partial_x,\partial_y)$, ${\bf v}_\perp = 
(v_x,v_y)$, ${\bf b}_\perp = (b_x,b_y)$. 
Furthermore, we ignore $v_z$ and $b_z$, in comparison with ${\bf v}_\perp$ and 
${\bf b}_\perp$, respectively (see below).  In addition to introducing anisotropy, for a non-zero $B_0$, there should be competition between the propagating Alfv\'en waves and the nonlinear terms in Eqs.~(\ref{eqmhd1a}) and (\ref{eqmhd2a}). The interplay between this competition and the anisotropy controls the ensuing scaling behaviour, as we show below.  It is evident from Eqs.~(\ref{eqmhd1a}) and (\ref{eqmhd2a}) that as $B_0$ increases, the nonlinear cascades are progressively weakened relative to the strength of the propagating modes. It is thus convenient to introduce a phenomenological dimensionless parameter
\begin{equation}
 M=\frac{B_0^2}{s^2},
\end{equation}
where $s$ is the typical magnitude of $v,b$ in the inertial range~\cite{foot22}. Depending upon the magnitude
of $B_0$, there are three possible physically distinct regimes: $M \ll 1,\,M\sim {\cal O}(1)$ and $M\gg 1$.

For small $M\ll 1$, the nonlinear terms dominate in the inertial range and the system becomes effectively isotropic. Unsurprisingly, the kinetic and magnetic energy spectra should then scale as $k^{-5/3}$ in the inertial range, in accordance with the K41 prediction.

 For a stronger $B_0$, when $M\sim {\cal O}(1)$ the linear propagating terms and the nonlinear terms in Eqs.~(\ref{eqmhd1a}) and (\ref{eqmhd2a}) are comparable in the inertial range, known as  strong turbulence.  We then balance 
\begin{equation}
 ({\bf v}_\perp\cdot {\boldsymbol\nabla}_\perp) {\bf v}_\perp \sim 
B_0\partial_z {\bf b}_\perp \implies 2a-1=-\xi +y.\label{anisobal1}
\end{equation}
Next, we balance
 the in-plane nonlinear terms in Eq.~(\ref{eqmhd2a}):
\begin{equation}
 ({\bf v}_\perp\cdot {\boldsymbol\nabla}_\perp) {\bf v}_\perp \sim ({\bf b}_\perp \cdot {\boldsymbol \nabla}_\perp) {\bf b}_\perp \implies a=y.
\end{equation}
Notice that this automatically gives
\begin{equation}
 \frac{\partial {\bf b}_\perp}{\partial t}\sim B_0 \frac{\partial}{\partial z} 
{\bf v}_\perp.\label{anisobal2}
\end{equation}
Using the dispersion relation (\ref{alfdisp}) we find
\begin{equation}
 \tilde z = \xi\implies a=y
\end{equation}
eventually.  Based on the physical arguments enunciated above, we continue to impose scale-independence of the magnetic or kinetic energy flux in the inertial range independent of $B_0$, which yields
\begin{equation}
 2y=\tilde z = 2a.
 \end{equation}
Therefore, we get
\begin{equation}
 \tilde z=\frac{2}{3}=\xi,\;a=\frac{1}{3}=y,
\end{equation}
 see, e.g., Ref.~\cite{gold1}.
Notice that the dynamic exponent $\tilde z$ is unchanged from its value from 
the isotropic case ($B_0=0$). Also, $\tilde z=\xi$ keeps (\ref{alfdisp}) unchanged under rescaling.

Enumeration of the scaling of the energy spectra requires extending the logic 
outlined above to anisotropic situation. Since $a=y$, we already expect 
identical scaling by the kinetic and magnetic energy spectra in the inertial 
range. We define Fourier transforms
\begin{eqnarray}
 {\bf s}_\perp ({\bf k}_\perp,k_\parallel,t)&\sim& \int {\bf s}_\perp ({\bf 
x}_\perp,z,t)d^2x_\perp dz\nonumber \\ &\sim& l_\perp^{a+2+\xi}\sim k_\perp^{-a-2-\xi}, 
\label{anisofour}
\end{eqnarray}
where $s=v$ or $b$. Now define
\begin{equation}
 \langle {\bf s}_\perp ({\bf k},t)\cdot {\bf s}_\perp ({\bf k'},t)\rangle= F_a (k) 
\delta ({\bf k}_\perp+{\bf k'}_\perp)\delta (k_\parallel + k_\parallel'),
\end{equation}
where $F_a(k)$ is related to $E_a(k)$ (see below).
Under scaling (\ref{resc-ani}), we obtain
\begin{equation}
 F_s (k_\perp,k_\parallel)\sim k_\perp^{-2a-2-\xi}.
\end{equation}
{ We can now use $k_\parallel\sim k_\perp^\xi$ and obtain the one-dimensional energy spectra as follows:
\begin{eqnarray}
 &&E_v(k_\perp)\sim E_b(k_\perp)\sim k_\perp^{-5/3},\\
 &&E_v(k_\parallel)\sim E_b(k_\parallel) \sim k_\parallel^{-2} \label{epara}
\end{eqnarray}
in the inertial range;  see, e.g., Ref.~\cite{gold1}. Results (\ref{epara}) can be obtained as follow: we note that the scaling ${\bf v}_\perp\sim l_\perp^a\sim l_\parallel^{a/\xi},\,{\bf b}_\perp\sim l_\perp^y\sim l_\parallel^{y/\xi}$. Scaling of one dimensional spectra $E_a(k_\parallel)$ follows from the equality
\begin{equation}
 E_{tot,s}=\int E_s(k_\parallel)\,dk_\parallel = \int E_s(k_\perp)\,dk_\perp,
\end{equation}
where $s=v,b$ and subscript $tot$ implies total energy (kinetic or magnetic). Then dimensionally,
\begin{equation}
 E_s(k_\parallel)\sim \left[\frac{ E_s(k_\perp) dk_\perp}{dk_\parallel}\right].
\end{equation}
This gives $E_v(k_\parallel)\sim k_\parallel^{-2a/\xi -1},\,E_b(k_\parallel)\sim k_\parallel^{-2y/\xi -1}$, giving (\ref{epara}) with $a=y=1/3,\xi=2/3$.
Thus, both $E_v(k_\perp)$ and $E_b(k_\perp)$ scale with $k_\perp$ 
according to the K41 result, but $E_v(k_\parallel)$ and $E_b(k_\parallel)$ scale {\em differently} with $k_\parallel$.}

We note that the result $\xi=2/3$ can be interpreted as singular renormalisation of $B_0$ in the 
long wavelength limit: we write the Alfv\'en wave term
\begin{equation}
 B_0 k_\parallel\sim B_0k_\perp^{2/3}\sim B_0(k_\perp) k_\perp,
\end{equation}
with $B_0 (k_\perp)\sim k_\perp^{-1/3}$. This is reminiscent of the result in 
Ref.~\cite{abjkb-jstat}.

We now provide {\em aposteriori} justification for neglecting $v_z,\,b_z$ the 
$z$-components of $\bf v$ and $\bf b$ in the above analysis. Notice that under 
scaling (\ref{resc-ani})
\begin{equation}
 {\bf s}_\perp \sim l_\perp^{1/3},\;\;s_z\sim l_\perp^0.
\end{equation}
The latter scaling essentially follows by demanding that different nonlinear terms involving $s_z$ and ${\bf s}_\perp$ in (\ref{eqmhd1a}) or (\ref{eqmhd2a}) scale in the same way. 
Thus, in the long wavelength limit $l_\perp\rightarrow \infty$, ${\bf s}_\perp 
\equiv (s_x,s_y)\gg s_z$. Hence, $E_v$ and $E_b$ are dominated by ${\bf 
v}_\perp$ and ${\bf b}_\perp$ in the long wavelength limit. This justifies our 
neglecting $v_z,\,b_z$ in the above analysis. 

We further expect the scaling behaviour to change substantially for much stronger $B_0$, i.e., with $M\gg 1$ for which the balances (in a scaling sense) used in (\ref{anisobal1}) and (\ref{anisobal2}) should breakdown, with the linear Alfv\'en wave terms dominating over the nonlinear terms in Eqs.~(\ref{eqmhd1a}) and (\ref{eqmhd2a}) even in the inertial range. For simplicity we do not distinguish between $s_z$ and ${\bf s}_\perp$. We note that in the limit of a very large $B_0$, the nonlinear terms should be suppressed. This in turn should lead to suppression of the energy fluxes, both kinetic and magnetic. In the limit of very large $M$, we express the energy fluxes $\epsilon_s$ phenomenologically (in a dimensional/scaling sense) as 
\begin{equation}
\epsilon_s \sim \frac{\partial s^2}{\partial t}\frac{1}{M}\label{anisoflux1}
\end{equation}
to the leading order in $1/M$; $s=v,b$. Imposing scale-independence of the fluxes then yields
\begin{equation}
 \tilde z=4a=4y.\label{kraich1}
\end{equation}
If we ignore anisotropy and consider $\xi=1$, then the dispersion relation (\ref{alfdisp}) yields
\begin{equation}
 \tilde z=1.
\end{equation}
This together with (\ref{kraich1}) gives
\begin{equation}
 a=y=\frac{1}{4}.
\end{equation}
Proceeding as above, this implies
\begin{equation}
 E_{v,b}(k)\sim k^{-3/2}.\label{ik1}
\end{equation}
which is the well-known IK spectra~\cite{ik}. %{\cred It has been argued that if there is a strong scale separation between the forcing scales and the inertial range, the magnetic field modes in the forcing scales can effectively act as background mean magnetic fields~\cite{pouquet1}. In that case, one is expected to observed the IK spectra given by (\ref{ik1}); see e.g., 
%Ref.~\cite{anupam}.}

The main criticism of the IK prediction is that despite having $M\gg 1$, anisotropy is ignored, 
which is not physically acceptable. We will now discuss how the scaling analysis 
for $M\gg 1$ may be affected by anisotropy.  We first note that while for $M\gg 1$ 
nonlinear terms are expected to be suppressed, dispersion relation (\ref{alfdisp}) in fact 
suggests that this suppression is ineffective for $k_\parallel^2\ll k_\perp^2$.
To account for this anisotropic suppression of the fluxes,  we phenomenologically 
modify (\ref{anisoflux1}) for the flux to (in a dimensional/scaling sense)
\begin{equation}
 \epsilon_s\sim \frac{\partial s^2}{\partial t}\frac{l_\parallel^2}{Ml_\perp^2},
\end{equation}
valid for $k_\perp^2\lesssim k_\parallel^2$. Now demanding scale-independence of the fluxes and using relevant time-scale $\sim l_\parallel$, we find 
(in a scaling sense)
\begin{equation}
 s\sim l_\perp^{1/2}l_\parallel^{-1/4},\label{anisok2}
\end{equation}
$s=v,\,b$.
We now assume that for $M\gg 1$, the nonlinear interactions leading to energy cascades predominantly take place 
only in the $xy$ plane (i.e., in the plane normal to $B_0 \hat z$) for very strong $B_0$
~\cite{ng1},  since the propagating Alfv\'en wave terms dominate along $\hat z$-directions. This implies that the 
kinetic and magnetic energy spectra are {\em solely} functions of $k_\perp$. This together with
(\ref{anisok2}) gives
\begin{equation}
 E_{v,b}(k_\perp,k_\parallel)\sim k_\perp^{-2}k_\parallel^{-1/2},
\end{equation}
corresponding to {\em weak turbulence limit}; see Refs.~\cite{ng1,perez1,gal2,gal3}.

\subsection{ 3D Hall MHD}
We now study scaling in Hall MHD (HMHD), first the isotropic case, then the corresponding
anisotropic one.

\subsubsection{Isotropic 3D HMHD}
\label{hmhd11}

In 3D Hall MHD (HMHD), one generalises the ordinary 3D MHD equations by 
including the Hall contribution in the form of the Ohm's law:
\begin{equation}
 {\bf E +v\times b}- \frac{\bf J\times b}{\rho_e}=\mu{\bf J},
\end{equation}
where $\rho_e$ is the electron charge density~\cite{galtier,galtier1}. This 
generalises Eq.~(\ref{eqmhd2}) to
\begin{eqnarray}
  \frac{\partial {\bf b}}{\partial t} &=&  {\boldsymbol \nabla}\times ({\bf v}\times {\bf b})\nonumber \\ &-&d_I 
{\boldsymbol\nabla} \times [({\boldsymbol\nabla}\times {\bf b})
\times {\bf b}] 
+ \mu\nabla^2 {\bf b} + {\bf f}_b, \label{eqhmhd}
\end{eqnarray}
where $d_I$ is the ion inertial length~\cite{galtier,galtier1}. We 
consider a vanishing mean magnetic field, i.e., $\langle {\bf b}\rangle 
=0$. Velocity $\bf v$ continues to obey 
Eq.~(\ref{eqmhd1}). Total energy $E$ remains a conserved quantity in HMHD in its ideal or inviscid limit~\cite{galtier1,hall-cons}. We ignore helicity for simplicity. % For reasons identical to those explained in Sec.~\ref{mhd1}, there are no 
%renormalisations for $\lambda_1,\,\lambda_2$. 

Evidently, for length scales $l\gg d_I$   the $d_I$-term is 
irrelevant compared to the first term on the rhs of (\ref{eqhmhd}) for a sufficiently large, and the 
scaling behaviour for ordinary isotropic 3D MHD ensues. In the opposite limit, the 
$d_I$-term is important. This range of scales is called the dispersion 
range~\cite{cho}; this does not exist for ordinary 3D MHD. We focus on the 
latter case, for which it suffices to 
ignore the $\lambda_1$-term~\cite{foot1}. While there is no symmetry principle 
that prohibits renormalisation of $d_I$ in a perturbative RG framework, 
considering $d_I\sim {\cal O}(1)$ and hence $l\lesssim {\cal O}(1)$, any 
perturbative corrections to $d_I$ stemming from the dispersion range should be 
``small'' and hence ignored in what 
follows below. 

We use scaling {\em ans\"atze} as defined by (\ref{resc1}) and (\ref{bsc}). Due to the 
rather different forms of the nonlinear terms in Eqs.~(\ref{eqmhd1}) and 
(\ref{eqhmhd}), same scaling of $\bf v$ and $\bf b$ is no longer expected. 
 In the dispersion range, $d_I{\boldsymbol\nabla} \times 
[({\boldsymbol\nabla}\times {\bf b})
\times {\bf b}]$ is the dominant nonlinear term in the rhs of (\ref{eqhmhd}). The dynamics of $\bf b$ is essentially controlled by the $d_I$-nonlinear term in the dispersion range.
We then balance
\begin{equation}
 \frac{\partial {\bf b}}{\partial t}\sim {\boldsymbol\nabla} \times 
[({\boldsymbol\nabla}\times {\bf b})
\times {\bf b}] \implies y=2-\tilde z.
\end{equation}
We continue to use $a=1-\tilde z$. Between the kinetic and magnetic 
fluxes, which flux is to be assumed to be scale-independent is crucial. We 
notice that demanding scale-independent magnetic flux yields
\begin{equation}
 2y=\tilde z\implies y=\frac{2}{3},\;\tilde z=\frac{4}{3}.
\end{equation}
This, however, gives $a=-1/3 <0$, which is clearly unphysical. We therefore 
discard this. In contrast, scale-independence of the kinetic energy flux yields
\begin{equation}
 2a=\tilde z\implies a=\frac{1}{3},\;\tilde z=\frac{2}{3},\; 
y=\frac{4}{3}.\label{hmhdscal}
\end{equation}
The scaling exponents (\ref{hmhdscal}) keep the whole of (\ref{eqhmhd}) 
scale-invariant, but breaks the scale-invariance of (\ref{eqmhd1}). More 
importantly, what are the physical dynamic exponents for $\bf v$ and $\bf b$ 
here that control the renormalisation of $\nu$ and $\mu$? Imposition of the 
scale-invariance of the kinetic energy flux ensures that the kinematic 
viscosity $\nu$ indeed picks up a scale-dependence $\sim l^{2-\tilde z}$. Since 
this choice of $\tilde z$ does not keep the magnetic flux scale-independent, we 
cannot say the same for the magnetic viscosity $\mu$, leaving the question of 
the physical dynamic exponent for $\bf b$ unresolved. This can however be 
settled by allowing for {\em two} different dynamic exponents $\tilde z_v$ and 
$\tilde z_b$, respectively for $\bf v$ and $\bf b$, such that 
scale-independence can be imposed on each of the kinetic and magnetic energy 
fluxes separately. This automatically yields
\begin{eqnarray}
 2a&=&\tilde z_v\implies a=\frac{1}{3},\,\tilde z_v = \frac{2}{3},\nonumber \\
 2y&=&\tilde z_b \implies y =\frac{2}{3},\,\tilde z_b = \frac{4}{3}. 
\label{exphmhd}
\end{eqnarray}
Thus, we obtain {\em weak dynamic scaling}~\cite{weak}.  That scale-independence of the kinetic and magnetic energy fluxes should imply two dynamic exponents $\tilde z_v$ and $\tilde z_b$ is in agreement with 
Ref.~\cite{chandra}. To our knowledge, there 
has been no 
systematic measurements of the time-scales of $\bf v$- and $\bf 
b$-fluctuations. This may however be measured in numerically, e.g., by 
calculating time-dependent correlation functions of $\bf v$ and $\bf b$ in 
pseudospectral methods~\cite{pseudo}.  With $\tilde z_b>\tilde z_v$, magnetic
fluctuations are longer lived. Hence, at sufficiently long time scales larger than $1/k^{\tilde z_v}$ but smaller than $1/k^{\tilde z_b}$, the 
$\bf v$-fluctuations die out, effectively making the $\bf b$-fluctuations 
autonomous. On the other hand, at shorter time scales, the magnetic field
fluctuations will appear frozen in time, and $\bf v$ effectively fluctuates in a given
background of spatially nonuniform but frozen in time $\bf b$.

Following the logic outlined in Sec.~\ref{mhd1} we can now obtain the scaling 
of the kinetic and magnetic energy spectra valid over length scales smaller 
than $d_I$. We find
\begin{equation}
 E_v(k)\sim k^{-1-2a}\sim k^{-5/3},\;E_b(k)\sim k^{-1-2y} \sim k^{-7/3}.
\end{equation}
Thus in this length-scales the magnetic energy spectrum is distinctly steeper 
than that kinetic energy spectra. Lastly, in the inertial range with length scale
$l\gg d_I$, unsurprisingly the scaling of $E_v(k)$ and $E_b(k)$ are identical to those in 3D MHD.

\subsubsection{Anisotropy effects}\label{hmhdaniso}

We now study the effects of spatial anisotropy brought in by a mean magnetic 
field $B_0$, assumed to be along the $z$-direction. Equation~(\ref{eqhmhd}) now 
generalises to
\begin{eqnarray}
  \frac{\partial {\bf b}}{\partial t} &+& \lambda_1({\bf v}\cdot 
{\boldsymbol\nabla}) {\bf 
b} = d_I({\bf b}\cdot {\boldsymbol\nabla}) {\bf v} -d_I 
{\boldsymbol\nabla} \times [({\boldsymbol\nabla}\times {\bf b})
\times {\bf b}] \nonumber \\&-&d_I B_0 \partial_z {\boldsymbol\nabla}\times 
{\bf b}
+ \mu\nabla^2 {\bf b} + {\bf f}_b. \label{eqhmhda}
\end{eqnarray}
Velocity $\bf v$ follows Eq.~(\ref{eqmhd1a}). Similar to the Alfv\'en waves, 
Eqs.~(\ref{eqmhd1a}) and (\ref{eqhmhda}) have circularly polarised whistler and 
cyclotron modes having dispersion of the form
\begin{equation}
 \omega \propto k k_\parallel
\end{equation}
for a large enough $d_I$~\cite{galtier1}. 

 We notice that with increasing $B_0$, the nonlinear terms in Eq.~(\ref{eqhmhda}) are progressively suppressed. Thus,  as in 3D anisotropic ordinary MHD, we expect different scaling behaviour for small or large $B_0$. We begin considering the situation when the linear and the nonlinear terms balance. This is the direct analogue of the strong limit of 3D anisotropic ordinary MHD turbulence. Given our results obtained 
in Sec.~\ref{mhda1} and Sec.~\ref{hmhd11}, we anticipate both anisotropy and 
weak dynamic scaling for length scales smaller than $d_I$ with $B_0\neq 0$. We 
proceed by using the scaling {\em ans\"atze} defined in (\ref{resc-ani}). Similar to 
Sec.~\ref{mhda1}, we ignore $v_z$ and $b_z$ in what follows below. First we 
balance
\begin{eqnarray}
 \frac{\partial {\bf b}_\perp}{\partial t}&\sim& d_I 
{\boldsymbol\nabla} \times [({\boldsymbol\nabla}\times {\bf b}_\perp)
\times {\bf b}_\perp]\nonumber \\ &\sim& \partial_z {\boldsymbol\nabla}\times 
{\bf b}_\perp \nonumber \\&&\implies y=2-\tilde z,\;\tilde 
z=1+\xi.\label{hmhdani0}
\end{eqnarray}
 Here we have implicitly assumed that $\xi <1$, and hence $k_\perp$ dominates over $k_\parallel$ in the dispersion regime. For strong $B_0$ following the logic developed above and
from Eq.~(\ref{eqmhd1})
\begin{eqnarray}
 &&({\bf v}_\perp\cdot {\boldsymbol\nabla}_\perp) {\bf v}_\perp \sim ({\bf v}_\perp\cdot {\boldsymbol\nabla}_\perp) 
{\bf v}_\perp \implies a=1-\tilde z, \label{hmhdani1}\\
&&\partial_t {\bf v}_\perp \sim B_0 \partial_z {\bf b}_\perp \implies a-\tilde 
z = y-\xi.\label{hmhdani2}
\end{eqnarray}
This is the analogue of the strong limit of 3D anisotropic MHD.
Assuming scale-independent kinetic energy flux, we obtain by using 
(\ref{hmhdani1})
\begin{equation}
 a=\frac{1}{3},\,\tilde z=\frac{2}{3},\,y=\frac{4}{3}.
\end{equation}
This unexpectedly gives $\xi=\tilde z-1 <0$, which is clearly unphysical. On 
the other hand, if we use (\ref{hmhdani2}) we get
\begin{equation}
 \xi = \tilde z - a + y =\frac{5}{3} >1\implies \tilde z>2,
\end{equation}
which is rather unexpected.
Inspired by our scaling analysis for isotropic HMHD, we try to resolve this by 
assuming two dynamic exponents $\tilde z_v$ and $\tilde z_b$, respectively, for 
$\bf v$ and $\bf b$. As in the isotropic case, we obtain $\tilde z_v$ by 
imposing scale-independence of the kinetic energy spectrum. We find
\begin{equation}
 \tilde z_v = \frac{2}{3},\,a=\frac{1}{3}.
\end{equation}
On the other hand, using (\ref{hmhdani0}) together with the condition of 
scale-independent magnetic energy spectrum, we get
\begin{equation}
 y=\frac{2}{3},\,\tilde z_b = \frac{4}{3},\,\xi=\frac{1}{3}.
\end{equation}
Thus, weak dynamic scaling is obtained. Further $\xi>0$ justifies our 
neglecting $v_z$ and $b_z$ in the above analysis; see discussions in 
Sec.~\ref{mhda1}. The existence of two dynamics exponents $\tilde z_v$ and 
$\tilde z_b$ is actually consistent with the original idea of 
Chandrasekhar~\cite{chandra}, which ensures scale-independence of the both 
kinetic and magnetic energy flux. As in isotropic 3D HMHD, $\tilde z_b>\tilde z_v$, implying 
magnetic fields to fluctuate independent of the velocity fields for sufficiently large time scales. Similar to its isotropic analogue, it would 
be interesting to verify this numerically.
It is now straight forward to obtain the scaling of the energy spectra. We 
obtain
\begin{eqnarray}
 E_v(k_\perp)&\sim& k_\perp^{-5/3},\\
 E_b(k_\perp)&\sim& k_\perp^{-7/3};\label{ebkhmhd}
\end{eqnarray}
 see also Ref.~\cite{gal3}. Analogously, we find $E_v\sim k_\parallel^{-3},\,E_b\sim k_\parallel^{-5}$.
Thus, in the dispersion range both $E_v(k)$ and $E_b(k)$ scale with $k_\perp$ in ways same their respective scaling with $k$ in the isotropic case, where as their scaling with $k_\parallel$ are markedly different. 

 For $B_0$ very large, for which the $d_I$-nonlinear term in (\ref{eqhmhda}) is strongly suppressed, the scaling of $E_b(k_\perp)$ is expected to change from (\ref{ebkhmhd}).
Several other possibilities for the scaling of $E_b(k_\perp)$ can then exist. For instance, for strong $B_0$ if we assume that the propagating mode sets sets the dynamic exponent $\tilde z_b$ and $k_\parallel$ and $k_\perp$ scale the same way, then $\tilde z_b =2$.  The condition of scale-independence of the magnetic flux yields $2y=\tilde z_b$, giving $y=1$. This then yields $E_b(k_\perp)\sim k_\perp^{-3}$, assuming the energy cascade is confined to the plane normal to ${\bf B}_0$; see, e.g.,~\cite{gal-prx}. 

 If we now account for spatial anisotropy (should be important for strong $B_0$) and phenomenologically express the scale-independence of the magnetic energy flux as
\begin{equation}
 \epsilon_b \sim \frac{\partial b^2}{\partial t}\frac{l_\parallel^2}{M_bl_\perp^2}\sim l^0,
\end{equation}
then $4y=\tilde z_b$, where $M_b=B_0^2/b^2$ with $b$ being the typical magnitude of the magnetic fields in the dispersion regime. Noting that the time-scale $\tau\sim l_\perp l_\parallel$, being controlled by the propagating mode, we have the anisotropic scaling of ${\bf b}$:
\begin{equation}
 {\bf b}\sim l_\perp^{3/4}l_\parallel^{-1/4}.
\end{equation}
This in turn gives 
\begin{equation}
 E_b(k)\sim k_\perp^{-5/2}k_\parallel^{-1/2},
\end{equation}
see Refs.~\cite{gal3,gal-prx}.

\subsection{3D electron MHD}

We now analyse the scaling behaviour of 3D EMHD - first the isotropic case, 
then the anisotropic version.

\subsubsection{Isotropic 3D EMHD}\label{emhdsec}

We first study the scaling in 3D isotropic electron MHD (EMHD).  
The EMHD equation for 
the magnetic field is in the absence of any mean magnetic 
field is~\cite{emhd1,emhd2}
\begin{eqnarray}
 \frac{\partial}{\partial t} ({\bf b} &-& \lambda_e^2 \nabla^2 {\bf b}) = -g 
{\boldsymbol\nabla} \times [({\boldsymbol\nabla}\times {\bf b}) \times ({\bf 
b-\lambda_e^2 \nabla^2  b})] \nonumber \\ &+&\mu \nabla^2 {\bf b} - \frac{\nu_e 
c^2}{\omega^2_{pe}} \nabla^4 {\bf b} + {\bf f}_b. \label{emhdeq}
\end{eqnarray}
Here, $\nu_ec^2/\omega^2_{pe}$ is a hyperviscosity. Although there are no 
symmetry arguments that prevent renormalisation 
of the coupling $g$, we ignore such issues here considering the fact that 
Eq.~(\ref{emhdeq}) is expected to be valid for sufficiently small scales for 
which fluctuation corrections should be small.  We ignore fluctuations in the density for simplicity.

We introduce the following scaling:
\begin{equation}
 {\bf x}\rightarrow l {\bf x},\,\,t\rightarrow l^{\tilde z} t,\,\, {\bf b} 
\rightarrow l^a {\bf },\label{rescemhd}
\end{equation}
For $\lambda_e^2/l^2\ll 1$, the $g$-nonlinear term in (\ref{emhdeq}) reduces 
to the standard Hall term in (\ref{eqhmhd}). We therefore focus on the opposite 
limit $\lambda_e^2/l^2\gg 1$ and ignore the fourth order hyperviscosity term in 
(\ref{emhdeq}). Equation~(\ref{emhdeq}) then reduces to
\begin{equation}
 \frac{\partial }{\partial t} \lambda_e^2 \nabla^2 {\bf b} = g {\boldsymbol 
\nabla}\times [\nabla^2 {\bf b} \times ({\boldsymbol\nabla}\times {\bf b})] + 
\mu \nabla^2 {\bf b} + {\bf f}_b. \label{emhdeq1} 
\end{equation}
Noting that EMHD description applies in small scales,  demanding scale-invariance we balance 
\begin{equation}
 \frac{\partial }{\partial t}  \nabla^2 {\bf b}\sim {\boldsymbol 
\nabla}\times [\nabla^2 {\bf b} \times ({\boldsymbol\nabla}\times {\bf b})]
\end{equation}
in the steady state, giving $y=2-\tilde z$. Constancy of the magnetic energy flux then implies
\begin{equation}
 2y=\tilde z \implies y=\frac{2}{3},\;\tilde z=\frac{4}{3},
\end{equation}
unchanged from isotropic HMHD. It is now straightforward to obtain
\begin{equation}
 E_b(k)\sim k^{-7/3},
\end{equation}
valid for length scales appropriate for EMHD, and identical to the scaling of $E_b(k)$ in the dispersion
range of 3D HMHD.

\subsubsection{Anisotropic effects}\label{emhdaniso}

In the presence of a mean magnetic field $B_0$ along the $z$-direction, 
(\ref{emhdeq1}) takes the form 
\begin{eqnarray}
  \frac{\partial {\bf b}}{\partial t} &&  = -g 
{\boldsymbol\nabla} \times [({\boldsymbol\nabla}\times {\bf b})
\times {\bf b}] \nonumber \\&+&g B_0 \partial_z {\boldsymbol\nabla}\times 
{\bf b}
+ \mu\nabla^2 {\bf b} + {\bf f}_b. \label{emhdeq11}
\end{eqnarray}
This leads to a dispersion (ignoring the viscous term)
\begin{equation}
 \omega \sim kk_\parallel.
\end{equation}
Now, introduce scaling 
\begin{equation}
{\bf x}\rightarrow l_\perp {\bf 
x}, {z}\rightarrow l_\parallel { z},\, t\rightarrow l_\perp^{\tilde 
z},\,{\bf b}\rightarrow l_\perp^y {\bf b}, 
\label{scemhd-ani}
\end{equation}
where $l_\perp$ is a length scale in the $xy$-plane.
As before, we 
further set length scale along the $z$-axis $l_\parallel \sim l_\perp^\xi$ that 
controls the relative scaling 
between the $xy$-plane and the $z$-axis; for $\xi\neq 1$, the system is 
anisotropic. It is now easy to extract the scaling exponents by 
directly following the logic outlined for 3D anisotropic HMHD.  We study the strong $B_0$ case when the linear and the nonlinear terms balance in the dispersion range. We find 
\begin{equation}
 E_b(k_\perp) \sim k_\perp^{-7/3},\;E_b(k_\parallel)\sim k_\parallel^{-5},\,\xi=\frac{1}{3},\,\tilde 
z_b=\frac{2}{3}.\label{emhdscaani}
\end{equation}
It is not a surprise that the above scaling in (\ref{emhdscaani}) is identical 
to the scaling obtained for $E_b(k)$, given the similarity between 
(\ref{emhdeq1}) and (\ref{eqhmhda}) with ${\bf v}=0$. Our results are actually 
quite close to those found in other studies. For 
instance, Refs.~\cite{cho,cho1} indeed found the magnetic energy 
spectra to scale as $k_\perp^{-7/3}$; the anisotropy 
exponent $\xi=1/3$ and $\tilde z_b=4/3$. This gives credence to our scaling 
analysis. 

 Similar to anisotropic 3D HMHD, the magnetic energy spectrum $E_b(k)$ in 
anisotropic 3D EMHD can display scaling $k^{-3}$ and $k_\perp^{-5/2}k_\parallel^{-1/2}$ under similar conditions. 

\section{Summary and outlook}\label{summ}

We have here revisited the  scaling of the magnetic and kinetic energy 
spectra in the various regimes of incompressible 3D MHD by developing a scaling 
theory. We obtain $k^{-5/3}$ scaling for both the kinetic and magnetic energy 
spectra in ordinary isotropic 3D MHD.  We further discuss the possibility of $k^{-2}$ spectra in isotropic 3D MHD.  The scaling theory predicts that the nature of scaling of the energy spectra in the anisotropic 3D MHD can be diverse, depending upon the strength of the mean magnetic field $B_0$, a feature that persists in anisotropic 3D Hall MHD and anisotropic 3D EMHD as well. For instance, when the magnitude of $B_0$ is such that the linear Alfv\'en wave terms balance the nonlinear terms in the inertial range,   
the scaling of both the kinetic and magnetic energy spectra with respect to $k_\perp$ is still given by the K41 result, but 
takes a different power law when expressed in terms of $k_\parallel$.  This is associated with an anisotropy exponent $\xi=2/3$ that relates the scaling of $k_\parallel$ with $k_\perp$. For very large $B_0$, for which the linear Alfv\'en wave terms dominate over the nonlinear terms in the inertial range, we find $E_v(k_\perp,k_\parallel)\sim E_b(k_\perp,k_\parallel)\sim k_\perp^{-2}k_\parallel^{-1/2}$. The scaling 
analysis also yields the IK scaling for strong $B_0$ if the spatial anisotropy is ignored. Interestingly, however, if one uses the 
scale-dependent version of $B_0$ in the IK scaling, one immediately gets back the 
K41 result.  These are in agreement with the existing results.
Independent of any spatial isotropy, we always get the same dynamic exponent for 
$\bf v$ and $\bf b$ in ordinary MHD, corresponding to strong dynamic scaling. 
High resolution numerical studies should complement the scaling results, settle the controversies surrounding scaling in 3D MHD.

For HMHD, we predict that the 
scaling of $E_b(k)$ in the dispersion range should be steeper than that for 
$E_v(k)$. This holds with or without a mean magnetic field. More interestingly, 
we predict two different dynamic exponents for $\bf v$ and $\bf b$. Since we find 
$\tilde z_b>\tilde z_v$, $\bf v$-fluctuations decay much faster than the $\bf 
b$-fluctuations. As a result, $\bf b$ fluctuations effectively appear as frozen 
fields in the dynamics of $\bf v$ over length scales belonging to the dispersion 
range, where as for sufficiently large time-scales, the dynamics of $\bf b$ 
fluctuations should be independent of the $\bf v$-fluctuations in the same 
regime, and hence reduces to 3D EMHD. This conclusion remains true whether or 
not there is a mean magnetic field. For 3D EMHD, the predictions from our 
analysis for $\bf b$ agrees with the same in 3D HMHD, which are again observed 
in relevant numerical studies~\cite{cho,cho1}. Our scaling analysis predicts
 $-7/3$ scaling for $E_b(k)$ in isotropic 3D EMHD and in the dispersion regime of isotropic 3D HMHD. In the corresponding anisotropic cases, $E_b(k_\perp)$ still scales as $k_\perp^{-7/3}$, but $E_b(k_\parallel)$ scales as $k_\parallel^{-5}$ with respect to $k_\parallel$.  We also identify an anisotropy exponent $-1/3$, different from its value in 3D anisotropic MHD. Notice that the the anisotropy exponent in 3D 
HMHD is half of that in 3D ordinary MHD. This means that the anisotropic 
effects and effective two-dimensionalisation is stronger in HMHD than ordinary 
MHD. In the limit of very strong magnetic $B_0$, we obtain $E_b(k_\perp,k_\parallel)\sim k_\perp^{-5/2}k_\parallel^{-1/2}$, which is the analogue of weak MHD turbulence in 3D HMHD. Scaling of the magnetic energy spectrum in 3D EMHD is found to be same as in 3D HMHD. In this context, we note that a recent study on table-top laser plasma 
revealed a $k^{-7/3}$ for high wavevectors at late times, indicative of an EMHD 
or HMHD like behaviour~\cite{nat-comm}.
It may be noted that Refs.~\cite{goldreich97,boldy1} predicted somewhat different scaling for the energy spectra. It will be interesting to see how our scaling approach may be extended or modified appropriately to account for the results in Refs.~\cite{goldreich97,boldy1}.

It is now well-accepted  that the universal properties of  fully developed 
turbulence - fluid or MHD - cannot be characterised by the two-point 
correlation functions (equivalently by the energy spectra) alone. Instead, one needs 
to calculate a hierarchy of {\em multiscaling} 
exponents for different order structure 
functions (including the two point ones)~\cite{frisch,cho,abmhd,debarghya}. Our 
scaling analysis is of course not adequate to 
capture multiscaling. Nonetheless,  different 
{\em multiscaling universality classes} of 3D MHD should be associated to different 
scaling regimes (e.g., ordinary MHD or HMHD) of MHD turbulence elucidated here. Thus, our 
scaling analysis should be helpful in delineating or classifying the possible universal 
multiscaling properties of MHD turbulence. 

The arguments behind our scaling analysis are sufficiently general, and should 
be applicable to a wider range of systems. Indeed, it will be interesting to 
apply these to related systems, e.g., compressible turbulence, rotating turbulence, turbulence in a 
binary fluid above and below the miscibility transition point, two-dimensional 
fluid and MHD turbulence. We hope our work will trigger new studies for these 
systems along the lines developed here.

\section{Acknowledgement}
We thank M. S. Janaki for useful suggestions and critical comments on the 
manuscript. One of us (A.B.) thanks the Alexander von Humboldt Stiftung, 
Germany for 
partial 
financial support 
through the Research Group Linkage Programme (2016).


\begin{thebibliography}{99}
\bibitem{multif} Detailed experiments and numerical studies indicate small 
correction to the K41 result. We ignore this issue here.
\bibitem{flux1} We ignore intermittency of local energy dissipation that is intimately connected to the multiscaling phenomena.
\bibitem{foot22} By not distinguishing between $v$ and $b$ in the definition of $M$, we implicitly assume $v\sim b$ (in a scaling sense), giving $a=y$ that is borne out by the results obtained below.
\bibitem{foot1} We essentially focus on intermediate length-scales much larger 
than the dissipative scales, but smaller than the scales at which ordinary MHD 
prevails.
\end{thebibliography}
\end{document}